\documentclass[twocolumn,amsmath,amssymb,aps,floatfix, longbibliography,superscriptaddress]{revtex4-1}
\usepackage{graphicx}
\usepackage{dcolumn}
\usepackage{bm}
\usepackage{amssymb}

\begin{document}


\title{System-spanning dynamically jammed region in response to impact of cornstarch and water suspensions}

\author{Benjamin Allen}
\affiliation{Department of Mechanical Engineering and Materials Science, Yale University, New Haven, CT 06520}
\affiliation{School of Natural Sciences, University of California, Merced, California 95343}
\author{Benjamin Sokol}
\affiliation{Department of Mechanical Engineering and Materials Science, Yale University, New Haven, CT 06520}
\author{Shomeek Mukhopadhyay}
\affiliation{Department of Mechanical Engineering and Materials Science, Yale University, New Haven, CT 06520}
\author{Rijan Maharjan}
\affiliation{Department of Mechanical Engineering and Materials Science, Yale University, New Haven, CT 06520}
\author{Eric Brown}
\affiliation{Department of Mechanical Engineering and Materials Science, Yale University, New Haven, CT 06520}
\affiliation{School of Natural Sciences, University of California, Merced, California 95343}

\date{\today}

\begin{abstract}
We experimentally characterize the impact response of concentrated suspensions of cornstarch and water.  We  hypothesize that the dynamically jammed region that  propagates ahead of the impactor is responsible for the strong stress response to impact when it spans between solid boundaries.    Using surface imaging and particle tracking at the boundary opposite the impactor, we observed that a visible structure and particle flow at the boundary occur with a delay after impact.  We show the delay time  is about the same time as the the strong stress response, confirming that the strong stress response results from deformation of the dynamically jammed structure once it spans between the impactor and a solid boundary.  A characterization of this strong stress response is reported in a companion paper \cite{MMASB17}.  We also elaborate on the structure of the dynamically jammed region  once it spans from the impactor to a solid boundary.  
We observed particle flow in the outer part of the dynamically jammed region at the bottom boundary,  with a net transverse displacement of up to about 5\% of the impactor displacement, indicating shear at the boundary.  Direct imaging of the surface of the outer part of the dynamically jammed region reveals  a change in surface structure that appears the same as the result of dilation in other cornstarch suspensions.  Imaging also reveals cracks, like a brittle solid.
These observations suggest the dynamically jammed structure can temporarily support stress according to an effective modulus, like a soil or dense granular material, along a network of frictional contacts between the impactor and solid boundary.  
\end{abstract}



\maketitle

Discontinuous Shear Thickening (DST) suspensions  exhibit a remarkable effect in which the suspensions behave like typical liquids at low shear rates, but when sheared faster the resistance to flow can increase discontinuously with shear rate \cite{Ba89,BJ14}.
This effect has been observed in a large variety of concentrated suspensions of hard, non-attractive particles, and is inferred to be a general feature of such suspensions \cite{Ba89, BJ12, BJ14,BFOZMBDJ10}. DST suspensions also support large stresses under impact, one example of which is the ability of a person to run on the surface of a pool filled with a suspension of cornstarch and water  \cite{youtube_running, BJ14}.  The impact response of such fluids is of practical interest for impact protection gear because of their strong response during impact while remaining fluid and flexible otherwise \cite{LWW03,D3O}. The purpose of this paper is to characterize the  internal structure of the suspension that leads to the strong impact response, to aid in developing models.   A companion paper focuses on characterizing the stress response \cite{MMASB17}.


Recently it has been found that a `dynamically jammed' region forms ahead of the impactor in the fluid, which moves along with the impactor like a plug \cite{WJ12}.    The dynamically jammed region grows during the impact, with a front propagating away from the impactor \cite{WJ12,PJ14,HPJ16}.   There is a sharp velocity gradient at the front, which separates the dynamically jammed region from the surrounding fluid \cite{PJ14}.   In a two-dimensional dry granular experiment the front velocity and width diverge at the same critical packing fraction as the viscosity curve of DST suspensions  \cite{WRVJ13}.  While the dynamically jammed region is propagating in the bulk, it is argued to exhibit no significant change in packing fraction \cite{HPJ16}.  Although it is presumed that this dynamically jammed region transmits stress via solid-solid frictional contacts, evidence of this is lacking. 

When the dynamically jammed region reaches the boundary, the stress increases beyond the prediction of the added mass effect \cite{PJ14}.  In a companion paper to this one, we report  a phenomenological constitutive relation between stress and strain for the impact response \cite{MMASB17}.  We find that when the dynamically jammed region (as defined by its contribution to the added mass effect) propagates to the boundary, the stress increases rapidly beyond the added mass effect, reaching up to the order of  MPa \cite{MMASB17}.  It is not yet known what is responsible for the stress scale on the order of MPa, yet it is 3 orders of magnitude larger than steady state shear in rheometer experiments. In rheometer experiments, the scale of the maximum shear stress supported in the shear thickening regime is limited by surface tension at the suspension-air boundary in response to dilation, and transmitted via frictional contacts \cite{BJ12}.  Whether there is any similarity in the force transmission mechanism between steady state shear and impact response remains to be seen.

In this paper, we propose the hypothesis that the dynamically jammed region could support a compressive load  that is transmitted via frictional interactions across the system  when the dynamically jammed region spans from the impactor to a solid stationary boundary.  This assumes that the solid boundaries are much harder and have much more inertial mass than the fluid, so the relatively soft  dynamically jammed region will deform  as it crashes into the stationary solid boundary.  The system-spanning dynamically jammed region could then temporarily support a load as it deforms according to its effective stiffness, perhaps strong enough to support a person running on the surface.    

To test this hypothesis and characterize the structure of the dynamically jammed region, we perform impact experiments while imaging and tracking particles at the boundary.  In our experiments, the impactor is driven far enough  into  a suspension to see the dynamically jammed region interacting with the boundary,  in contrast to previous experiments which probed mainly the response of the bulk \cite{WJ12,WRVJ13, PJ14, HPJ16}, but not so close to the boundary  to be affected by short-range boundary effects (i.e.~within $\approx 3$ mm) \cite{LSZ10}. Our experiments are at $V_I$  faster than quasistatic compression, so that  dynamically jammed fronts can exist, but at speeds slow enough that inertial effects  \cite{Bagnold54,CB13} including added mass \cite{WJ12} and high Mach number effects \cite{LPWG10,POLMFH15, GKL17}) are negligible. This intermediate velocity regime is where the steady-state DST transition occurs (typically at flow velocities $\stackrel{<}{_\sim} 10^{-2}$ m/s in rheometers \cite{MB17}), but surprisingly, systematic force measurements have not yet been reported in this regime as far as we know.

The remainder of the paper is organized as follows.   The materials and methods  of  suspension impact experiments are explained in Sections \ref{sec:materials} and \ref{sec:methods}, respectively.  In Sec.~\ref{sec:visualization} we show images  at the boundary of the suspension  that reveal structural changes in the dynamically jammed region that appear to be a consequence of dilation. In Sec.~\ref{sec:displacement}, we present particle tracking measurements at the boundary to identify where particle flow and shear occurs.  In Sec.~\ref{sec:onset} we compare the timing of the stress increase with  that of the onset of motion of tracked particles at the boundary to confirm that the the stress increase is a consequence of the dynamically jammed region spanning between solid boundaries.

\section{Materials}
\label{sec:materials}

The suspensions were made of  cornstarch  purchased from Carolina Biological Supply, and tap water  near room temperature. Weight fractions $\phi$ for cornstarch and water were measured as the weight of the cornstarch divided  by the total weight of cornstarch plus water.   Weight fractions of cornstarch and water are very sensitive to histories of temperature and humidity \cite{SF44}, so different data sets taken with relative humidity ranging from 8\% to 54\% are not directly comparable.    To avoid misinterpretation from false comparisons, we do not report weight fractions for different experiments.  All samples nominally had weight fractions from 0.53 to 0.61, in a range where they all exhibited noticeable shear thickening when stirred by hand.   For data sets represented in a single plot, the experiments were taken over a short enough time period to have a humidity standard deviation of  6\%. Measurements were made at a temperature of $22.0\pm0.6^{\circ}$C. The density of the suspensions is $1200\pm20$ kg/m$^3$ \cite{MMASB17}.

  Samples  were initially mixed on a vortex mixer until  no dry powder chunks were  observed. Before each impact measurement, samples were additionally stirred by slicing through them at least 5 times with a spatula at velocities low enough to avoid significant cracking of the suspension and prevent large air bubbles from being trapped inside the suspension.   This additional stirring helps counter any systematic effects of settling  or compaction from previous experiments.  This procedure produced a level of reproducibility of $\pm 30\%$  in stress measurements,  equivalent to what we could achieve by making new samples  before each measurement.  If instead we did not stir between measurements or we forced air bubbles to get trapped in the suspension, the stress varied by around a factor of 2 from run to run.

\section{Methods}
\label{sec:methods}

\begin{figure}
\centering
\vspace{0.2in}
{\includegraphics[width=0.35 \textwidth]{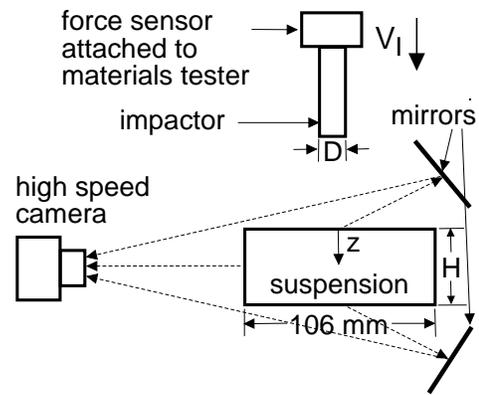}}
\caption{Schematic  of the experimental setup (side view).  Measurement are made of the mean normal stress $\tau$ on the impactor as a function of impactor depth $z$ and impactor  velocity $V_I$. This can be done simultaneously with imaging of the top, bottom and side boundaries  of the suspension. }
\label{fig:apparatus}
\end{figure}

We performed experiments to visualize the  boundary of the suspension to observe the dynamically jammed region, while simultaneously measuring forces  in response to impact.  We used a high-speed camera (Phantom M110) to image the suspension boundaries  through a transparent acrylic container with a square base, and containing a  suspension of cornstarch and water, as shown in Fig.~\ref{fig:apparatus}.   The top, bottom, and side  boundaries  can all be observed using mirrors.   In most experiments (unless otherwise stated), a cylindrical aluminum impactor of diameter $D = 12.7$ mm was pushed into a container with a square base of length $106$ mm, with the suspension filled to a height $H=42$ mm.  These   dimensions are such that the  region  of interest below the impactor is far from the sidewalls of the container.   The impactor surface unintentionally had a slight wedge shape, which was angled at  $4^{\circ}$ relative to the surface.   This can be seen to produce some asymmetry in displacements measured in Sec.~\ref{sec:displacement}. We used an Instron E-1000 dynamic materials tester to push the impactor into the fluid at constant velocity $V_I$, while measuring the normal force on the impactor as a function of  depth $z$ from the free surface of the suspension (downward positive).   The nominal relative position resolution within each run is 1 $\mu$m.  We define $z=0$ and time $t=0$ at the top surface of the suspension, with an uncertainty of 0.5 mm.   The impactor started at a height typically $5.0 \pm 0.5$ mm above the suspension surface  and was pushed to a final position typically within $10\%$ of the bottom of the container.  While the impactor had a set point constant impact velocity $V_I$,  it had to accelerate at the beginning and end of the test. This resulted in a standard deviation of the velocity of the impactor of typically 11\% while $z>0$.  We measured a mean normal stress $\tau$ on the impactor. Calibration of stress measurements and their results are explained in the companion paper \cite{MMASB17}.

\section{Results}

\subsection{Quasi two-dimensional  visualization}
\label{sec:visualization}

\begin{figure}
\centering
\vspace{0.2in}
{\includegraphics[width=0.475 \textwidth]{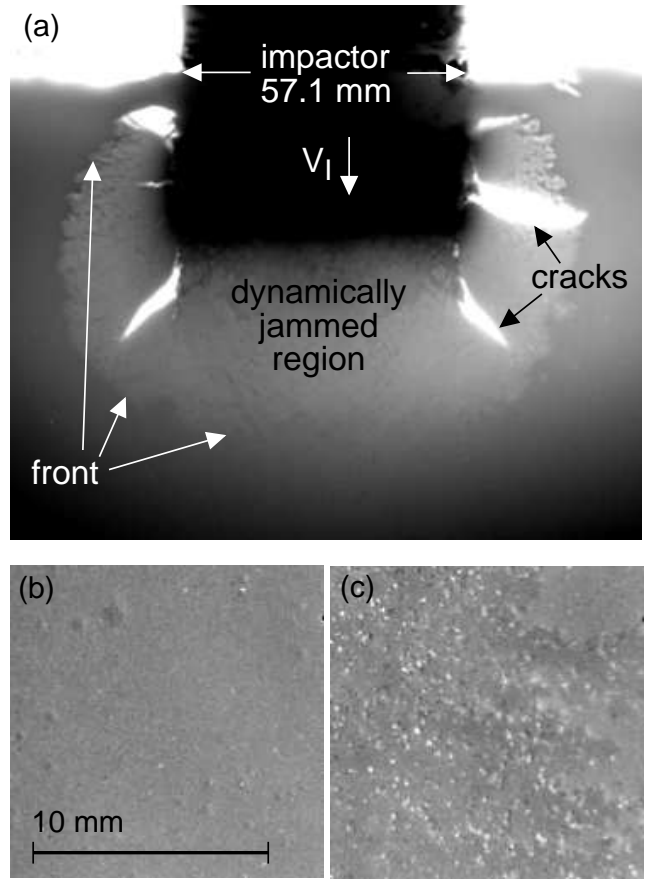}}

\caption{(a) A side view shows the  dynamically jammed region ahead of the impactor in a quasi-2-dimensional experiment.  The dynamically jammed region grows over time as its front propagates ahead of the impactor towards the bottom of the image.  Fractures inside this region suggest it has some characteristics of a brittle solid. (b) zoomed in view of the region ahead of the impactor in a similar experiment before impact. (c) same view as panel b, but after impact, so that the view is of the dynamically jammed region.  The surface appears rougher, which may be due to particles poking through the surface as a result of dilation.}
\label{fig:2Dfrontimage}
\end{figure}

To observe the growth of the dynamically jammed region, we first show a side view of a quasi two-dimensional version of the apparatus in Fig.~\ref{fig:2Dfrontimage}a.  This experiment is designed for qualitative observation only.    A solid rectangular impactor of width 57.1 mm and thickness 5 mm was pushed into a suspension at  $ V_I=100$ mm/s in an $8$ mm thick cell.  We observed a lighter-shaded region, indicative of some kind of change in surface structure, roughly semi-circular in shape, which first appears near the impactor and grows as its front propagates outward faster than the impactor.  This front propagation can be observed in Supplementary Video 1 which is played back at 0.1 times real time.  This light region appears to be the dynamically jammed region proposed by Waitukaitis \& Jaeger \cite{WJ12}, with propagation similar to that found by \cite{PJ14}.  The change in reflected light off the surface indicates a change in surface structure.  Cracks can be seen as bright spots due to the image being mostly backlit,  which have been observed previously at the surface of DST suspensions \cite{RMJKS13}.   In repeated experiments we observed that the cracks can appear in different places, preferentially starting at the sharp corners of the impactor or at its sides. The cracks never propagate outside of the  dynamically jammed region past the propagating front.  This indicates that the dynamically jammed region has some characteristics of a brittle solid, as proposed by Waitukaitus \& Jaeger \cite{WJ12} while the outer region remains fluid-like.   

The lighter-shaded portion of the dynamic jammed region in Fig.~\ref{fig:2Dfrontimage} appears  matte or rough to the naked eye.  The image and video appear very similar to what is seen when a suspension of cornstarch and water dilates under shear (Fig.~11 of \cite{BJ12}), tension (Fig.~3 of \cite{SBCB10}), or compression (Fig.~1b of \cite{RMJKS13}), or when a person steps on the wet sand at the beach that a wave has recently passed by.  Dilation is a common result when dense granular flows are sheared, in which the particles push around each other and the packing expands, while the voids between particles enlarge. The rough surface appearance is the result of particles poking through the liquid-air interface of the suspension at the visible surface, while the liquid retreats  into the interior to fill the larger voids opened by the dilating particle structure.  Cornstarch particles are too small to be seen individually by the naked eye, but  scatter reflected light in different directions like a rough surface \cite{BJ12,BZFMBDJ11}.  The lighter shading in Fig.~\ref{fig:2Dfrontimage}a is expected as the result of the light reflecting off the surface being mostly indirect, so more light is scattered back to the camera by the rough surface.  

To more clearly show the  change in surface structure,  we took a zoomed-in video of the lighter-shaded portion of the dynamically jammed region in a similar experiment.  Snapshots of a 13 mm square region are shown before the impact in Fig.~\ref{fig:2Dfrontimage}b, and after the impact in Fig.~\ref{fig:2Dfrontimage}c,  so that the latter view is of the dynamically jammed region.   The  surface of the dynamically jammed region scatters light more diffusively,  indicating a rougher surface.  Supplementary Video 2 shows a few square centimeter section at the wall in the lighter-shaded region behind the front before, during, and after impact, played back at 0.1 times real time.  Initially the quiescent surface has a uniform reflectivity, indicating it is smooth, except for some trapped bubbles air that are on the order of 1 mm in size (some can also be seen in Fig.~\ref{fig:2Dfrontimage}b), much larger than individual particles.  After flow starts (2 seconds into the video), the surface appears rough.  During this time, the macroscopic air bubbles disappear, likely as a result of dilation of the particle packing, and liquid can be seen to be drawn away from the wall, resulting in air pockets  between the suspension and the wall.  The rough appearance of the surface is expected if particles poke through the liquid-air interface as a result of dilation \cite{BJ12}.  Note that the presence of the container wall affects the  quantitative values of interfacial tensions, but  is not expected to prevent  the liquid from drawing away from the surface and being replaced by air (which could be drawn from nearby trapped air bubbles,  dissolved gas in the water, or cavitation).  After the impactor stops (7 seconds into the video), the dynamically jammed region is seen to retreat and the surface reflectivity becomes more uniform again.  Air bubbles are observed to reform  out of the  rough patches during this relaxation,  confirming that there was air  in between the suspension and the wall during impact in the rough-looking regions.  This surface change which appears to us to be a result of dilation is in contrast to the bulk propagation of the dynamically jammed region, where it is argued to exhibit no decrease in packing fraction \cite{HPJ16}.

 \subsection{Visualization at the boundary of a three-dimensional system}

  \begin{figure} 
\centering
\vspace{0.2in}
{\includegraphics[width=.475 \textwidth]{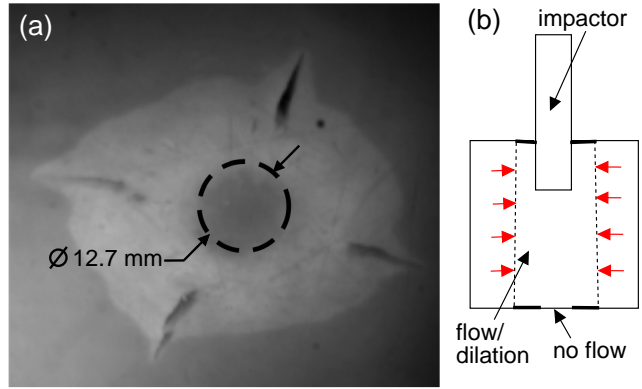}}
\vspace{0.2in}
\caption{A picture showing the dynamically jammed region at the bottom boundary of the suspension centered below the impactor.   The impactor outline is indicated by the dashed circle of diameter $D=12.7$ mm.   (b)  A schematic (not to scale)   identifying corresponding properties of the dynamically jammed region when it spans from the impactor to the opposite boundary.  A  dead zone with no particle flow is found on the bottom  boundary with roughly the diameter of the impactor.  A  surrounding cylindrical region exhibits particle flow, apparent dilation, and sometimes cracking.  Red arrows illustrate that a confining stress on the sides of the dynamically jammed region is required  to support a compressive load on the impactor via a system-spanning frictional contact network.
}
\label{fig:surfaceimages}
\end{figure}

 In order to see how this dynamically jammed region behaves in a 3-dimensional system, we imaged a front-lit suspension at the bottom boundary of the square-base container described in Sec.~\ref{sec:methods} with a high-speed camera at  up to 1000 frames per second.  
 An example is shown in Fig.~\ref{fig:surfaceimages}a  at the maximum penetration depth of  $z= 39.5$ mm (2.5 mm from the bottom boundary) for $V_I = 396$ mm/s.   We observed a localized change in intensity of reflected light on the bottom boundary, in a roughly  annular shape centered directly below the impactor, about 3 to 4 times the radius of the impactor.   Just outside the impactor radius, this structure appears similar to the observation in Fig.~\ref{fig:2Dfrontimage}, suggesting that this is the dynamically jammed region, and may be the result of dilation.    
 We observed cracks, which appear dark in this case because they do not extend through the entire suspension.  They only form if $\tau > 4\cdot 10^6$ Pa, but did not appear systematically. Moreover, the pattern is not always symmetric.
 Such a stress could correspond to the ultimate strength of  a solid-like material.   Finally, there is also a central circle about the same size as the impactor with a less noticeable intensity change indicating a different structure in the center.

  The time evolution of the bottom boundary image can be seen in Supplementary Video 3, which is played back at 0.05 times real time.  The dynamically jammed region  appears with a delay after impact, similar to the stress response \cite{MMASB17}.  It appears first directly below the impactor, and grows radially outward.  After the impactor stops, the outer edge of the lighter region gradually retreats as the suspension returns to its liquid-like state over about 1 s.

In addition to the  observations at the bottom boundary, we also observed  a similar change in reflected light intensity on the top surface near the impactor, in agreement with previous observations \cite{RMJKS13}.   At the sides of the container we observed no intensity change in the experiments whose data is presented in this paper and in the companion paper \cite{MMASB17}.   These observations suggest a columnar dynamically jammed structure spanning from the impactor down to the bottom boundary as  illustrated in  Fig.~\ref{fig:surfaceimages}b.   The column has a nearly circular cross-section at the bottom (Fig.~\ref{fig:surfaceimages}a),  and at the top \cite{RMJKS13}.  We do not have enough information to specify how the diameter of the column  varies with depth.


We also performed experiments where the impactor was closer to the sidewall as in Fig.~\ref{fig:2Dfrontimage}.  In this case we observed the same visual change at the sidewall as at the top and bottom boundaries.  This is  expected as the dynamically jammed region propagates not only below the impactor but also to the sides \cite{PJ14,HPJ16}.  In these cases we observed that the stress increased sharply with a delay, about the same time as it took to observe the dynamically jammed region at the side boundary,  but shorter than it took for the dynamically jammed region to reach the bottom when the impactor was further from the side walls.  This indicates that in such geometries the sidewalls may support the load, similar to other dense granular systems \cite{janssen1895}.  

 \subsection{Particle tracking}
\label{sec:displacement}

To determine how much compression and/or shear is in the dynamically jammed region, we included tracer particles in the suspension during the stress measurements shown in Fig.~2 of \cite{MMASB17}.   In those measurements there was a delay before the stress increased above a weak background, and reached up to the order of MPa.  The tracer particles were iron-oxide particles of diameter $a = 0.12 \pm 0.02$ mm. These are heavy and large enough to settle on the bottom boundary and be visible there.  On the other hand, they are still small enough to act as tracer particles as it should only take a shear stress of $\sim \rho g a \approx 5$ Pa to overcome friction with the base and move them.  This can be achieved even if a tiny fraction of the $\sim$ MPa normal stress on the impactor is transferred into shear stress at the bottom.    

We tracked particles using the ImageJ multitracker plugin.  To filter out false tracks,  usually due to the tracking algorithm picking up different particles at different locations but falsely identifying them as the same particle, we threw out tracked particles that  in one frame moved more than a  threshold distance,  which we varied from 0.25--0.60 mm from run to run.  While we set  this to ensure eliminating all false tracks,  it also resulted in throwing out some real particles corresponding to those that moved the furthest over the course of the experiment, so  the following results underestimate the number of particles that moved between 2-6 mm  over the course of the experiment. 

 \begin{figure}
\centering
\vspace{0.2in}
{\includegraphics[width=0.475 \textwidth]{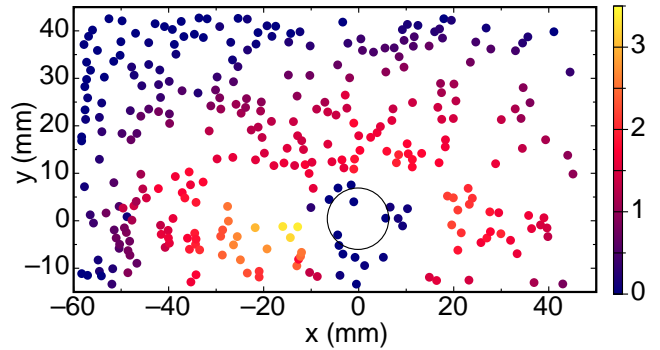}}
\vspace{0.2in}
\caption {(color online) A displacement field showing each particle tracked at the bottom boundary of the suspension for $V_I=114$ mm/s.  The color legend indicates the net radial displacement  $d_{r,n}$ of each tracked particle.   The circle  indicates the outline of the impactor. All  particles moved outward radially from the  axis of the impactor, which is used  as the origin  of the coordinate system.   A dead zone with no particle motion is observed directly below the impactor. 
} 
\label{fig:displacementfield}
\end{figure}

We calculated the  magnitude of the radial component of displacement $d_r$ of each particle as a function of time $t$.  From that we calculated the net radial displacement  $d_{r,n}$ measured from the beginning of the video (before impact) to the time the impactor reached its maximum  depth.  An example field map of the net radial displacement  $d_{r,n}$  is plotted in Fig.~\ref{fig:displacementfield} at $V_I=114$ mm/s for the same data set shown in Fig.~2 of \cite{MMASB17}, where the  maximum impactor depth was $z=34$ mm.  The  points are plotted at the starting position of each particle in the x-y plane on the bottom boundary,  where the point on the impactor axis is used  as the origin of the x-y plane. No significant negative values of $d_r$ were found,  indicating the particle flow was moving outward radially from the origin, as might  be  qualitatively expected in a fluid as material is displaced away from the impactor to create a circulating flow.   The field map of $d_{r,n}$ is close to, but not quite, radially symmetric. This asymmetry may be due to the slight wedge shape of the impactor which was angled at  $4^{\circ}$ relative to the surface for this set of experiments.

 \begin{figure}
\centering
\vspace{0.2in}
{\includegraphics[width=0.475 \textwidth]{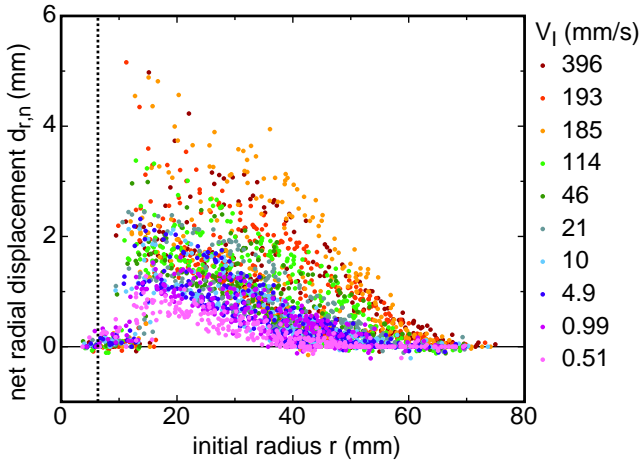}}
\vspace{0.2in}
\caption {(color online) The net radial displacement  $d_{r,n}$ of each tracked particle over the course of the impact as a function of its initial radius $r$.   Different  impact velocities $V_I$ are indicated in the legend.  The vertical line indicates the impactor radius. The dead zone at the center is similar at all impact velocities.  The gradient indicates about a 5\% compressive strain in the radial direction, and thus also shear within the system.
} 
\label{fig:displacement_radius}
\end{figure}

The same data for the net radial displacement  $d_{r,n}$ over  the course of the impact is plotted as a scatter plot in Fig.~\ref{fig:displacement_radius}.   A point is shown for each particle tracked as a function of the initial particle radius $r$, where $r$ is measured from the origin on the axis of the impactor.  Data for different $V_I$  are shown, and observed to have a  similar profile.  There is a large scatter in the data, despite the smoothness of the  variation  between neighboring points in the field map of Fig.~\ref{fig:displacementfield}.  Rather, the  scatter in Fig.~\ref{fig:displacement_radius}  is due to the radial asymmetry of the pattern in Fig.~\ref{fig:displacementfield}.   Note that our filtering method tends to cut out particles with $d_{r,n}$ ranging from  2-6 mm, so this plot may underestimate the  number of particles in this range of $d_{r,n}$. 

In Figs.~\ref{fig:displacementfield} and \ref{fig:displacement_radius}, a dead zone with no  significant particle motion is observed directly below the impactor with a radius of 10 to 15 mm, about twice the radius of the impactor.  The dead zone  is a feature that would not occur in a Newtonian fluid,  which  would only be expected to have an infinitesimal stagnation point at the center of the image.   Rather, such a dead zone is common in granular flows \cite{ETGZN13}. While the particles may not be moving noticeably in the dead zone, it is still expected that the contact forces fluctuate rapidly and irregularly due to the particle motion at the boundary of the dead zone, transmitting forces in a similar way as the flowing portion of the dynamically jammed region, rather than a static force distribution like statically jammed systems.  The sharp increase in  $d_{r,n}$  at the edge of the dead zone indicates a strong shear (shear strain is equal to the gradient of displacement) at the interface of the dead zone and the outer portion of the dynamically jammed region,  and a well-defined boundary between two flow regimes.  This shear profile is in contrast to previous measurements taken at times before  the  dynamically jammed region spans to the boundaries, where it was found that the low-shear plug-like region is relatively large compared to the region of strong shear, and the shear increases gradually and monotonically moving away from the center of the dynamically jammed region \cite{WJ12,PJ14,HPJ16}. It is unclear yet if this difference in profile is due to the particular geometry and boundary conditions of the different experiments, or is a consequence of the dynamically jammed region colliding with the solid boundary to generate shear-bands.

 \begin{figure}
\centering
\vspace{0.2in}
{\includegraphics[width=0.475 \textwidth]{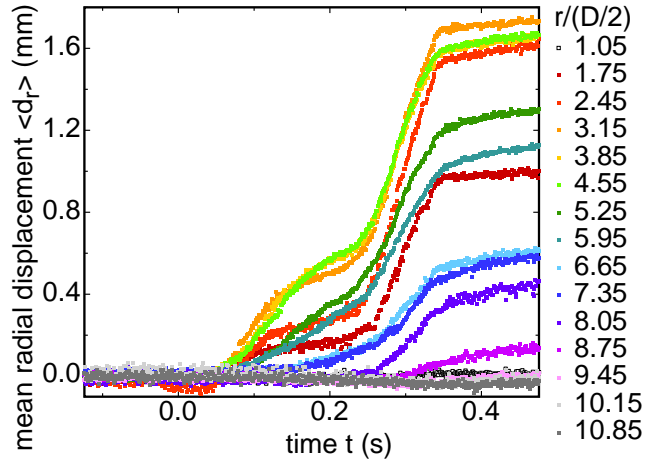}}
\vspace{0.2in}
\caption {(color online) The radial displacement $\langle d_r \rangle$  averaged over all tracked particles as a function of time $t$ at $V_I=114$ mm/s.   Particle displacements are averaged in bands of different initial radii $r$ shown in the key, in units of impactor radii.  Upper curves tend to correspond to smaller $r$ as long as they are outside of the dead zone ($R > 2D/2$).   A similar delay time is observed before any displacement at each radius.  The displacement levels off when the impactor comes to a stop ($t=0.36$ s).
} 
\label{fig:displacement_time}
\end{figure}

Figure \ref{fig:displacement_time} shows the time dependence of the radial displacement $\langle d_r\rangle$ for $V_I=114$ mm/s averaged over many particles in different ranges of initial $r$.   The bottom surface shown in Fig.~\ref{fig:displacementfield} is divided into concentric annuli of width $0.7D/2$, and displacements $d_r$ are averaged for all particles that started within each annulus.  Time series for each annulus are referred to in the legend by their mean radius $r$.   At each radius there is a delay before a significant increase in $\langle d_r \rangle$, similar to the delay in the stress response to impact in the same experiment \cite{MMASB17}.  In Fig.~\ref{fig:displacement_time}, it can be seen that all of the curves for $r < 8D/2$ increase above the background at the same time, so the delay time is independent of $r$.  For larger $r$, this delay time increases with $r$.  At the end of  each experiment, the impactor  decelerated rapidly to a stop.  For example in Fig.~\ref{fig:displacement_time}, this rapid stop is responsible for the kink seen in each curve at $t=0.36$ s.  At this time, the  load  on the impactor reached its peak value, the impactor velocity dropped to 50\% of its set point $V_I$, and the impactor was within 8\% of its final displacement.  After this rapid  impactor deceleration, the radial particle displacement $\langle d_r \rangle$  leveled off. This indicates minimal flow after the impactor stopped.

 The  net transverse displacement $\langle d_{r,n} \rangle$ just outside the dead zone was only about 5\% of the  displacement of the impactor after the  delay time, i.e.~after the particles started moving.  The strain in the radial direction can be calculated as the gradient in Fig.~\ref{fig:displacementfield}, which yields around 5\% compressive strain in the outer part of the dynamically jammed region, but much more significant local strain in shear bands at the boundary of the dead zone and near the wall.  Considering that there is compressive strain in both the vertical and radial directions indicates that the dynamically jammed region is not incompressible, and there must be shear strain in the system.  A stress transformation suggests the absolute maximum shear strain in a plane extending in the vertical and radial directions is half the larger of the principal strains, corresponding to a shear strain of 44\% when the impactor reaches its maximum penetration depth.
 
Compressive strain and shear can also be observed based on the relative motion seen at the side of the quasi-2-dimensional experiments shown in supplementary video 1, as in \cite{PJ14}.  For example, by tracking features in the central  region directly below the impactor, an average compressive  strain around of 30\% can be seen before the impactor stops.  On the sides of the impactor, shear can be observed as the impactor moves down while the cracks remain nearly still.  This implies an average shear strain on the sides of about 70\%.   While we can observe an average strain and shear at the boundaries, we do not have local information  about values of shear strain in the interior, which may vary around these averages.
 
By comparing  the displacement measurements from the particle tracking with the boundary visualization in Fig.~\ref{fig:surfaceimages}a,  we can  elaborate on the structure of the dynamically jammed region, as illustrated in Fig.~\ref{fig:surfaceimages}b.   The outer part of the dynamically jammed region exhibits particle flow and shear, as indicated by the particle tracking measurements in Fig.~\ref{fig:displacement_time}.   The shear in the dynamically jammed region can lead to dilation, which can draw liquid from the surface of the dynamically jammed region, which appears to be the case in Figs.~\ref{fig:2Dfrontimage} and \ref{fig:surfaceimages}a.  The suggestion of dilation in the dynamically jammed region when it spans between solid boundaries are in contrast to what is claimed for the dynamically  jammed region  before it spans between solid boundaries \cite{HPJ16}.  In the central portion of the dynamically jammed region, directly in front of the impactor on the opposite boundary and with radius $r \stackrel{<}{_\sim}0.7D$,  there is a different type of structure where there is  no particle flow, indicated by the particle tracking measurements in  Fig.~\ref{fig:displacement_time}.  
 

\subsection{Onset time comparison}
\label{sec:onset}

In this section we compare the onset time of the particle motion at the bottom boundary (Fig.~\ref{fig:displacement_time}) with that of the sharp increase in stress \cite{MMASB17}. This will test the hypothesis that this strong stress increase is the result of the deformation of the dynamically jammed region once it spans between solid boundaries.

 To compare timings between stress and particle displacement, the particle tracking measurements (Sec.~\ref{sec:displacement}) were taken simultaneously with the stress measurements in Fig.~2 of our companion paper \cite{MMASB17}. We calibrate the timing between the video and stress measurements by tracking a flag attached to the impactor.  We align the time at which the flag has moved 1 pixel in the video with the time that the impactor has moved  an equivalent distance.  

\begin{figure}
\centering
\vspace{0.2in}
{\includegraphics[width=0.475 \textwidth]{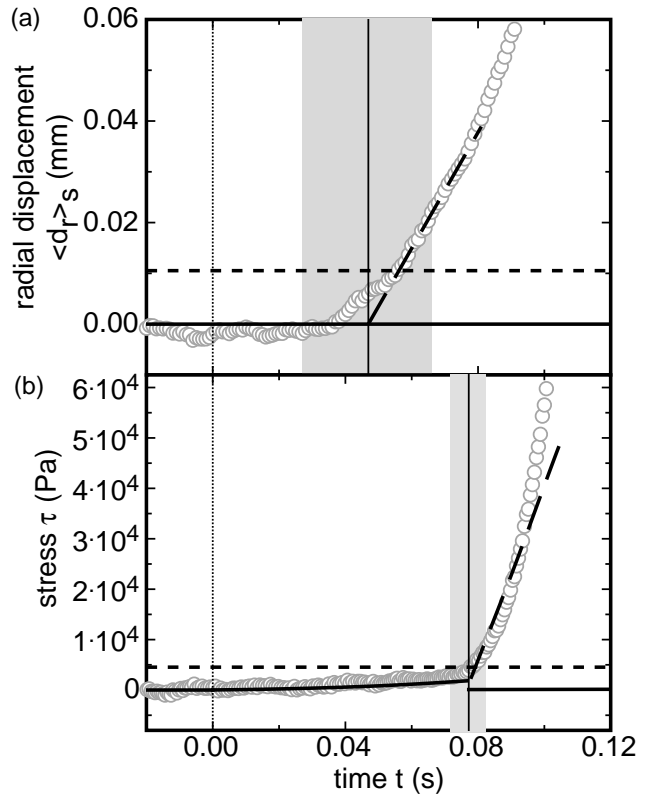}}
\vspace{0.2in}
\caption {Simultaneous time series of particle displacement and stress  for $V_I=114$ mm/s. (a) Smoothed mean radial displacement $\langle d_{r }\rangle_s$  of tracked particles. (b) Stress $\tau$.   Solid lines: background signals.  Short dashed lines:  threshold for fitting.  Long dashed lines: linear fit to the data above the  threshold, used to obtain onset time by extrapolating to the background. Vertical solid lines: onset times $T_d$ and $T_{\tau}$, respectively, of signal increases above the background, with the gray bands as uncertainties.  Vertical dotted line: the time of impact ($t=0$).  While both signals increase with a comparable delay after impact,  the stress increase occurs after particle displacement is observed at the bottom boundary.
} 
\label{fig:onset}
\end{figure}

To determine the onset time for particle tracking, we follow  as closely as possible the algorithm for finding the onset of the stress increase above the background in the companion paper \cite{MMASB17}.  We start with the time-dependent mean  radial particle displacement  $\langle d_r \rangle$ averaged over all particles at all radii from Sec.~\ref{sec:displacement}.   We smooth $\langle d_r \rangle$  uniformly over a range of $\pm0.5$ mm in $z$, just as we did for $\tau$ \cite{MMASB17} to obtain $\langle d_r \rangle_s$.  In Fig.~\ref{fig:onset} we show representative time series of the simultaneous particle displacement $\langle d_r \rangle_s$  (panel a) and stress $\tau$ (panel b) for $V_I=114$ mm/s.  The background from  buoyancy and the added mass effect is also shown as the solid line in panel b for the stress data \cite{MMASB17}.  

To identify the onset times, we fit each data set in Fig.~\ref{fig:onset}  after the signal first exceeds a threshold value \cite{MMASB17}.   The stress data from the simultaneous experiments is fit as a function of depth $z$ to Eq.~3 of the companion paper \cite{MMASB17} to include the background from added mass and buoyancy.
The particle displacement $\langle d_r \rangle_s$  is fit directly as a function of time, with a threshold value of $5\sigma_d$, where $\sigma_d$ is the standard deviation of $\langle d_r \rangle_s$  from the beginning of recorded data to $t=0$ when the impactor hits the surface. We linearly fit both $\langle d_r \rangle_s$ and $\tau$ starting from their  respective threshold values over a fixed range of  3 mm. This fit range is larger than the typically 1 mm range used in the companion paper\cite{MMASB17},  due to the larger noise in the particle tracking data.  Each fit is extrapolated to its respective background signal to obtain an onset time $T_d$ for particle displacement or $T_{\tau}$ for stress.   For particle displacement, the relevant background is the mean value of $\langle d_r \rangle_s$ for $t<0$.  For stress measurements, the fits and extrapolation were done as a function of depth $z$, since that was inferred to be the parameter that the force depends directly on \cite{MMASB17}.  The raw data table with both depth and time data was used as a lookup table to convert the depth at which the extrapolated fit reached the background signal to an onset time $T_{\tau}$.  Using the raw data table for this conversion eliminates most of the error from the variation in impact velocity $V_I$ over  the course of the experiment.  

The solid vertical lines shown in Fig.~\ref{fig:onset}  indicate the best estimate of the onset time $T_d$ or $T_{\tau}$ for each case.  The  gray bands are the error bars.  The uncertainty on $T_{\tau}$ comes from the uncertainty from the added mass effect, the possible range of the contribution of the background after $T_{\tau}$,  and the error  from the  extrapolation of the fit of $d\tau/dz$ to the point of intersection as explained in the companion paper \cite{MMASB17}. The uncertainty on $T_d$ relative to its zero only includes the errors of the fits of $\langle d_r\rangle_s$ propagated to the point of intersection.  We additionally include the relative error between the two timings based on the flag tracking on the error in $T_d$ only.  This relative uncertainty on $T_d$ includes the time between frames as a time resolution, plus the time it takes for the impactor to move the flag its first pixel distance to account for the limited spatial resolution of the video, plus an error propagated from the  0.5 mm error in position from smoothing.  The 0.5 mm absolute error on the position does  not contribute to a relative error  on the timing between the particle tracking and stress measurements  because it is the same  systematic error in both cases, so is not shown in Fig.~\ref{fig:onset}. The relative timing error is the largest error in Fig.~\ref{fig:onset}, and the total error is 40\%  of $T_d$ at $V_I=114$ mm/s.    Given this error, the delay before the onset of  tracked particle motion $T_d$  is still found to be smaller than the delay before the onset of stress $T_{\tau}$ by more than 1 standard deviation. 


\begin{figure}
\centering
\vspace{0.2in}
{\includegraphics[width=0.475 \textwidth]{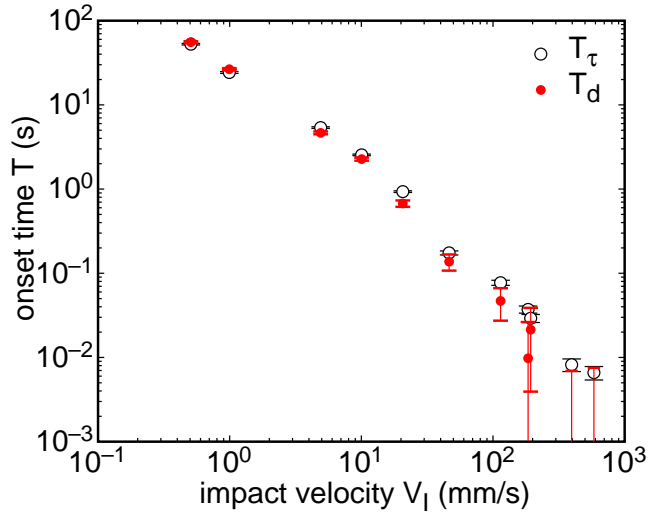}}
\vspace{0.2in}
\caption{Onset times $T_{\tau}$ for stress (open circles), and $T_d$ for particle displacement (closed circles), measured in simultaneous experiments, as a function of impact velocity $V_I$. The onset times are consistent with the stress  increase starting at the same time or shortly after the particle motion starts,  suggesting that the sharp stress increase at time $T_{\tau}$ requires a  signal first be transmitted from the impactor to the bottom boundary at time $T_d$.
} 
\label{fig:onset_vi}
\end{figure}

To obtain more statistics for this comparison, we plot $T_d$ and $T_{\tau}$ at different  impact velocities $V_I$ in Fig.~\ref{fig:onset_vi}. The errors shown are obtained the same way as those in Fig.~\ref{fig:onset}. The relative error between the two timings based on the flag tracking  is comparable to $T_d$ for larger $V_I$ as the  onset times  approach the flag timing resolution.  For  $V_I \ge 396$ mm/s where the absolute timing error is larger than $T_d$ we can only give an upper bound on $T_d$, indicated by the error bars overlapping with $T_d=0$  in Fig.~\ref{fig:onset_vi}.  The two  delay times $T_{\tau}$ and $T_d$ correspond  closely to each other  over the range measured, with a  root-mean-square difference between the onset times  of 14\% for $V_I \le 114$ mm/s where $T_d$ is resolvable,  while the onset times vary over 3 decades.  The stress increase usually occurs after particle flow is observed, with the non-overlap of the systematic error bars (i.e.~the resolution and the smoothing error) no more than 2 standard deviations of the random error (i.e.~the fit errors) when $T_{\tau}$ is smaller than $T_d$, while the non-overlap of the systematic errors is as large as 8 standard deviations of the random error when $T_{\tau}$ is larger than $T_d$.  These errors are statistically consistent with the stress increasing at time $T_{\tau}$ equal to or larger than  the onset of particle motion at time $T_d$.  This close correspondence in delay times, and usually slightly larger $T_{\tau}$ than $T_d$, indicates the large stress increase in the impact response first requires the propagation of a force signal from the impactor to the opposite boundary large enough to initiate particle displacement at the bottom.   This close correspondence between stress and particle tracking in Fig.~\ref{fig:onset_vi} also confirms that this front is the same as observed in quasi two-dimensional interface tracking \cite{PJ14}.   

The small difference in onset times could indicate a return travel time is required for the force signal to travel from the bottom of the container back up to the impactor where the  force sensor is located. If this were the case, the relatively small difference between $T_d$ and $T_{\tau}$ would  suggest the return travel is faster than that of the downward moving front,  which is reasonable since on the return trip the material is already in some way dynamically jammed and would be  expected to transmit force  more like a solid \cite{AK06}.   However,  given the large measurement errors compared to the difference between the two delay times, we do not have the resolution to calculate such a return travel  speed  with accuracy.   At best we can  put a  lower bound on the return speed that is at least a few times faster than the downward  signal propagation.

\section{Conclusions}

  On impact into a suspension, a dynamically jammed region appears in front of the impactor, and propagates ahead of the impactor (Fig.~\ref{fig:2Dfrontimage}).  This  is known to be responsible for the added mass contribution to impact response of suspensions \cite{WJ12}.   We showed that the  delay before stress response is consistent with or follows shortly after the onset of particle motion at the bottom boundary  (Figs.~\ref{fig:displacement_time}, \ref{fig:onset}, and \ref{fig:onset_vi}).     This demonstrates that the strong stress response to impact on the order of $10^6$ Pa \cite{MMASB17} requires deformation of the dynamically jammed region once it spans between solid boundaries.
  
 We observed two different sub-regions of the dynamically jammed region at the bottom boundary.  This includes a dead zone about the same size as the impactor cross-section with no particle flow in the central part of the dynamically jammed  region (Figs.~\ref{fig:surfaceimages}a, \ref{fig:displacementfield}, and \ref{fig:displacement_radius}).  In the outer part of the dynamically jammed region at the bottom boundary, we observed particle flow with a net displacement of up to 5\% of the impactor displacement (Fig.~\ref{fig:displacement_radius}), indicating shear within the dynamically jammed region.    We observe a change  in surface structure that  appears to be the same as the result of dilation in a dense granular suspension (Figs.~\ref{fig:2Dfrontimage}, \ref{fig:surfaceimages}a).  This  dilation is  in contrast to  the dynamically jammed region as it propagates through the bulk, where it is argued to exhibit no decrease in packing fraction \cite{HPJ16}. The outer part of the dynamically jammed region also cracks like a brittle solid.

The stress on the order of $10^6$ Pa \cite{MMASB17} reached in the impact response is so high that it implies that particles have been pushed together beyond the point where lubrication models break down, leading to effectively frictional interactions between neighboring particles where shear stress is proportional to normal stress  \cite{VG88, MMASB17}.  The frictional interactions,  fluid-like ability to flow, the appearance of dilation, and cracking, all suggest the dynamically jammed region behaves mechanically much like a soil or dense granular material.  
The observation of a stress proportional to deformation via an effective modulus is also similar to a soil or dense granular material \cite{MMASB17}.


 In soils and other dense granular systems stress is transmitted across the system via frictional interactions, and the scale is determined by the normal stress at the boundary \cite{LW69},  rather than being determined by an intrinsic constitutive rheology in terms of a local shear stress as a function of shear rate and packing fraction.   However, for the case of  impact, it is not yet clear what is the physical origin for the scale of the normal stress on the order of $10^6$ Pa.  
 
 It is interesting to compare and contrast this system with DST in steady-state shear, which is often assumed to be related to the impact response.  DST in steady-state shear is triggered by frustration of dilation by a confining stress at a boundary, along with force transmission between solid-solid frictional particle contacts \cite{BJ12, SMMD13, MSMD14, FMRKMLCHSI13, Heussinger13, RBH16, SGM17}.   In steady-state,  the structure has time to become well-developed so the stress distribution is more uniform through the suspension, and must be supported at all boundaries, so it is limited by the stiffness of the weakest element in series.   In most rheometer experiments, the weakest stiffness comes from surface tension at  the suspension-air  interface, which limits the  maximum normal stress that can be transmitted through the system to about $10^3$ Pa in steady state.   In  cases where the stress is not limited by the suspension-air interface, the weakest stiffness could be soft walls \cite{BJ12} or the  particle stiffness \cite{WB09, BJ12} -- the latter case has been observed  in steady state flows in simulations with periodic boundary conditions \cite{OH11, SMMD13, SGM17}, but not in hard-particle experiments.  In contrast, in transient impact, the dynamically jammed region does not have time to propagate to the side wall, and the stress does not have time to become uniform throughout the suspension.  Instead,  the sides of the columnar dynamically jammed region need support, as illustrated in Fig.~\ref{fig:surfaceimages}b.  The origin of the force that supports the sides of the dynamically jammed region is not yet known.
 
One possible force that appears in the transient but not the steady state comes from  flow through the pores between particles that open up during dilation in the transient.  The need to rapidly move liquid as the dynamically jammed region is dilating could introduce significant stresses on the dynamically jammed particle structure.   A rough estimate predicts a stress from this pore pressure on the scale of $\tau_p\approx \eta_l \alpha\Delta\phi V_I L/\kappa$, where the viscosity of the interstitial liquid is $\eta_l=9\times10^{-4}$ Pa$\cdot$s, the permeability $\kappa=(1-\phi^3)a^2/180\phi^2$, $\alpha$ is a dimensionless coefficient of order 1,  $L$ is the  width of the sheared region, and we interpret $\Delta\phi$ as the  change in weight fraction due to dilation from the initial value \cite{JVF16}.  If we assume $\alpha=4$ \cite{JVF16}, an estimate for a typical value of $\Delta\phi\approx 0.01$ in a dilating suspension, and $L\approx 1.5$ cm based on the size of the outer sheared region in Fig.~\ref{fig:surfaceimages}a at $V_I=396$ mm/s, then we obtain $\tau_p \approx 8$ MPa,  on the same order of magnitude as the  maximum stress observed in our companion paper \cite{MMASB17}.  It remains to be confirmed if this pore pressure is what sets the scale of the stress response to impact.

\section{Acknowledgements}

We thank Abe Clark, Bob Behringer, Scott Waitukaitis, Ivo Peters, and Heinrich Jaeger for discussions and for sharing their unpublished results. This work was supported by the NSF through grant DMR 1410157.

\section{Supplementary Videos}

Supplementary videos may be downloaded from:
\newline https://www.eng.yale.edu/brown/publications.html

%

\end{document}